\documentclass{article}
\usepackage{spconf}
\usepackage{amsmath,graphicx, xcolor, url}
\usepackage{cite,amsfonts,amssymb,psfrag,amsthm,paralist}

\usepackage{graphicx}
\usepackage{epstopdf}
\usepackage{rotating}
\usepackage{dsfont}
\usepackage{color}
\usepackage{tikz}
\usetikzlibrary{plotmarks}
\usepackage{pgfplots}
\pgfplotsset{compat=1.10}

\def\c{{\mathbf c}}
\def\x{{\mathbf x}}
\def\y{{\mathbf y}}
\def\n{{\mathbf n}}
\def\b{{\mathbf b}}

\def\S{{\mathbf S}}

\def\teachhash{{\mathbf h}_\text{T}}
\def\studhash{{\mathbf h}_\text{S}}
\def\pavg{{P_{\text{avg}}}}

\usepackage{flushend}

\def\BibTeX{{\rm B\kern-.05em{\sc i\kern-.025em b}\kern-.08em T\kern-.1667em\lower.7ex\hbox{E}\kern-.125emX}}

\title{MULTI-HOP DEEP JOINT SOURCE-CHANNEL CODING WITH DEEP HASH DISTILLATION FOR SEMANTICALLY ALIGNED IMAGE RECOVERY}

\name{Didrik Bergstr{\"o}m$^1$ \quad Deniz G{\"u}nd{\"u}z$^2$ \quad Onur G{\"u}nl{\"u}$^{3,\, 1}$}

\address{$^1$Information Theory and Security Laboratory (ITSL), Link{\"o}ping University, Sweden \\
	$^2$Department of Electrical and Electronic Engineering, Imperial College London, UK\\
    $^3$Lehrstuhl für Nachrichtentechnik, Technische Universität Dortmund, Germany}

\begin{document}
	\topmargin=0mm

	\maketitle

	\begin{abstract}
		We consider image transmission via deep joint source-channel coding (DeepJSCC) over multi-hop additive white Gaussian noise (AWGN) channels by training a DeepJSCC encoder-decoder pair with a pre-trained deep hash distillation (DHD) module to semantically cluster images, facilitating security-oriented applications through enhanced semantic consistency and improving the perceptual reconstruction quality. We train the DeepJSCC module to both reduce mean squared error (MSE) and minimize cosine distance between DHD hashes of source and reconstructed images. Significantly improved perceptual quality as a result of semantic alignment is illustrated for different multi-hop settings, for which classical DeepJSCC may suffer from noise accumulation, measured by the learned perceptual image patch similarity (LPIPS) metric.
	\end{abstract}

	\begin{keywords}
		Joint source-channel coding, DeepJSCC, multi-hop relaying, deep hash distillation, semantic communications.
	\end{keywords}
    
	\section{Introduction}\label{sec:intro} 
    Conventional communication systems work by first removing redundancy in data (source coding) and then adding structured redundancy against channel noise (channel coding). While Shannon's separation theorem shows that this approach is asymptotically optimal, it is known to be suboptimal for practical block lengths~\cite{JSCCVectorQuant}. Optimal performance is achieved by directly mapping the input signal to the channel codeword, called joint source-channel coding (JSCC). JSCC is a highly challenging problem due to the large dimensionality and lack of structure~\cite{JSCCFundamentals}. Until recently, no practically feasible and competitive designs have been known for general sources and channels. Benefiting from recent advances in deep learning methods, DeepJSCC~\cite{DeepJSCCWirelessImage} outperforms state-of-the-art separation-based baselines. Moreover, since its introduction, DeepJSCC has been extended to adapt to channel SNR and bandwidth~\cite{DeepJSCC-1++}, along with other modalities; see, e.g.,~\cite{DeepJSCCVideo, LowLatdeepJSCC}.
    
    An important limitation of DeepJSCC is noise accumulation in multi-hop relaying settings, where consecutive transmissions through noisy channels significantly degrade the quality of the reconstructed image, in terms of both distortion and perceptual quality~\cite{hybridDJSCCmulti-hop}. Continuous-amplitude nature of DeepJSCC prevents complete noise removal, achieved through channel coding in conventional systems. Distorted data also makes traditional cryptographic authentication infeasible, as modern methods assume data is reconstructed perfectly. Recent research has applied deep neural networks (DNNs) to \emph{hashing} for image retrieval. Deep hash distillation (DHD) method~\cite{DeepHashDist} trains a DNN that displays a notion of semantic understanding of images through unsupervised learning. DHD applies semantic clustering by generating ``fingerprints" corresponding to the semantic content of a source image, and these fingerprints are similar when they are generated from images with similar semantic content. Combining this property with DeepJSCC is an instance of \emph{semantic communication}, emphasizing the communication of the underlying ``meaning" of the data~\cite{Semantic} or the computation-relevant parts~\cite{OurDeepRDFC}. 

    In this paper, we propose a new architecture that incorporates DHD into the DeepJSCC framework, which can be considered a form of ``semantic clustering" that allows relays to mitigate semantic shifts caused by channel noise. We extend simplified point-to-point DeepJSCC-DHD designs in~\cite{DeepJSCCandDHDmaster} to multi-hop decode-and-forward (DF) relaying, by adding a semantic alignment mechanism that mitigates noise accumulation and enables security-oriented applications in a noisy domain. We also investigate the impact of channel output quantization on semantic alignment in multi-hop quantize-and-forward (QF) relaying. Our results show that the proposed approach, through semantic clustering, can mitigate noise accumulation while improving perceptual quality, highlighting its potential for deployment in practical multi-hop communication systems. Moreover, unlike training for perception, our design explicitly aligns the reconstructed image to a frozen DHD hash, conserving the semantic meaning of the source at the destination and enabling security-oriented applications.
    
	\section{Problem Formulation}\label{sec:probform}
    Consider a source image \(\S\!\in\! \mathbb{R}^{C \times H \times W}\), where \(C\), \(H\), and \(W\) denote the number of color channels, height, and width, respectively. We aim to wirelessly transmit \(\S\) to its destination via \(r\) relay nodes \(\{\text{R}_1,\dots,\text{R}_r\}\), where adjacent nodes are connected by complex additive white Gaussian noise (AWGN) channels with additive noise terms \(\n_i\), where \(i=1,\!\dots,r\!+\!1\). The AWGN components \(\n_i\) are considered to be mutually independent and identically distributed, i.e., we have \(\n_1\sim\!\n_2\!\sim\! \dots\!\sim\n_{r+1}\!\sim\!\mathcal{CN}(0, \sigma^2\mathbf{I}_k)\) for \(r\!+\!1\) hops, where \(k\) denotes the number of complex channel symbols. This simplified model allows us to gain fundamental insights from the experimental results. 
    
    We define the bandwidth ratio as $\rho \triangleq \frac{k}{CHW}$ channel symbols/pixel, and denote the signal-to-noise ratio (SNR) as \(\text{SNR}\triangleq10\log_{10}(1/\sigma^2)\)~dB.
    
    We measure the reconstruction quality of reconstructed images \(\widehat{\S}\) with the average peak SNR (PSNR), defined as
    \begin{align}\label{eq:psnr_def}
        \text{PSNR} = -10\cdot\log_{10}\left(\frac{\|\S-\widehat{\S}\|^2_2}{CHW}\right) \text{ dB}
    \end{align}
    and the \emph{perceptual} quality with LPIPS \cite{perceptualMetrics}, which has been shown to better align with human perception. We next introduce our baseline protocols.
    
    \subsection{Decode-and-Forward (DF) Protocol}
    DF multi-hop relaying can be considered as a sequence of point-to-point transmissions. Here, an encoder \(f_{i}\) transforms a source image \(\S\) to the channel input \(\x_{i} \in \mathbb{C}^k\) with an average power constraint $\frac{1}{k}\|\x_i\|^2 \leq \pavg := 1$. The first relay \(\text{R}_1\) receives the channel output \(\y_{i} = \x_{i} + \n_{i}\), and decodes \(\y_i\) to an intermediary representation \(\widetilde{\S}_i\in\mathbb{R}^{C \times H \times W}\) using a decoder \(d_i\). It then re-encodes \(\widetilde{\S}_i\) to \(\x_{i+1}\) using an encoder \(f_{i+1}\) which is transmitted to \(\text{R}_{i+1}\), and so on.

    \subsection{Quantize-and-Forward (QF) Protocol}
    Quantization operations are less complex than decoding operations, which in principle means simpler circuitry and lower total energy consumption~\cite[Chapter~14.5]{GerhardLectureNotes}. These properties make a QF relaying protocol well-suited for, e.g., relays in remote locations on the edges of core networks or low-latency satellite communications, as they will mainly quantize the received signal and relay it forward through noiseless pipelines obtained by using error correcting codes for each hop, akin to the setup in~\cite{hybridDJSCCmulti-hop}.
    
    Consider an encoder-decoder pair \((f_{Q}, d_{Q})\) that is adapted to transmit through an AWGN channel with a fixed SNR. We want to quantize the channel output \(\y\) observed at relay \(\text{R}_1\) to a bit sequence \(\b\) and forward it through a noiseless pipeline (e.g., perfect channel coding) to the destination decoder (\(\text{R}_2 \rightarrow  \cdots \rightarrow \text{R}_r\)). The sequence \(\b\) is then dequantized (mapped) to \(\widehat{\y}\), which is finally used by the decoder \(d_{Q}\) to reconstruct the quantized image \(\widehat{\S}\). We consider the ``naive quantization" approach in~\cite{hybridDJSCCmulti-hop} as a baseline, where we partition \(\y \in \mathbb{R}^{2k}\) into \(2k/N_Q\) blocks. Denote \(N_Q \geq 1\) as the number of real-valued elements of \(\y\) per block, and quantize each block with \(N_Qb\) bits. We compute the centers of the \(2^{N_Qb}\) codewords with the K-means algorithm~\cite{LloydKMeans}, and assume that the codebook is available at the relay \(\text{R}_1\) and the destination. The rate of the vector quantizer is expressed as bits per pixel (bpp), and we compute this rate as \(I = \frac{2k N_Qb}{N_Q HW}\).
    
    \section{Proposed Scheme}\label{sec:deepjscc_dhd}
    
    We propose a scheme that combines DeepJSCC and DHD to leverage the semantic clustering capabilities of the DHD module for enhanced semantic alignment between source and reconstruction images for multi-hop relaying systems.
    
    \def\DHDmodule{{\mathcal{H}}}
    \subsection{Deep Hash Distillation (DHD)}\label{subsec:dhd}
    A DHD module \(\DHDmodule(\cdot) \triangleq H_{\theta}(E_{\theta}(\cdot))\) consists of two parts 
    \begin{align*}
        E_\theta(\S) &:  \S \in[0,1]^{C \times H\times W} \rightarrow \mathbf{z} \in \mathbb{R}^{N_E}, \\
        H_\theta(\mathbf{z})& : \mathbf{z} \in \mathbb{R}^{N_E} \rightarrow \mathbf{h} \in (-1,1)^{N_H}
    \end{align*}
    where \(E_{\theta}(\cdot)\) is a pre-trained encoder that takes a source image \(\S\) and outputs a feature vector~\(\mathbf{z}\), and \(H_{\theta}(\cdot)\) is a fully connected (FC) hash function with \textit{tanh} activation that takes \(\mathbf{z}\) as input and outputs a hash vector~\(\mathbf{h}\) of length \(N_H\).

    The training procedure is to generate two transformed source images \(\S_\text{T}\) and \(\S_\text{S}\) from a source image \(\S\) in such a way that \(\S_\text{T}\) is less transformed/distorted than \(\S_\text{S}\). This mimics a knowledge distillation approach where the module transfers hashing knowledge from ``simple" to ``difficult" transformations~\cite{DeepHashDist}. Using the transformed source images as inputs, the DHD module outputs corresponding continuous-valued hashes \(\teachhash\) and \(\studhash\) that are inputs to the self-distilled hashing loss defined as
    \begin{align}\label{eq:sdh_loss}
        \mathcal{L}_{\text{SdH}}(\teachhash, \studhash) \triangleq 1-\mathcal{S}(\teachhash, \studhash)
    \end{align}
    where \(\mathcal{S}(\teachhash,\studhash) \triangleq \frac{\teachhash \cdot \studhash}{|\teachhash\|\studhash|}\) is the cosine similarity function.
    
    A key goal of DHD is to quantize its hash output \(\mathbf{h}\) to binary bits during inference using the \verb|sign| operation, where (\ref{eq:sdh_loss}) minimizes the cosine distance between continuous hashes to minimize the Hamming distance between the quantized binary hashes~\cite{DeepHashDist}. The quantization error is minimized by
    \begin{align}\label{eq:bce-q_loss}
        \mathcal{L}_{\text{bce-}Q}(\teachhash) \triangleq \frac{1}{K} \sum_{k=1}^{N_H}(H_b(b_k^+,g_k^+)+H_b(b_k^-,g_k^-))
    \end{align}
    where \(H_b(u,v) \triangleq -u\log_2(v)-(1-u)\log_2(1-v)\) is the binary cross entropy; \(g_k^{+}\) and \(g_k^{-}\) are maximum likelihood estimates of the \(k\)th hash element via Gaussian distributions \(g(\mathbf{h}_k) = \exp\left(\frac{-(\mathbf{h}_k-m)^2}{2\sigma_{g}^2}\right)\) with respective means \(m\!=\!+1\) and \(m\!=\!-1\); and \(b_k^{+}\!=\!\frac{1}{2}(\text{sign}(\mathbf{h}_k)+1)\), \(b_k^{-}\!=\!1-b_k^{+}\) denoting the binary likelihood labels.

    A proxy-based representation learning approach to hashing is introduced in~\cite{DeepHashDist} to, among other things, imbue the hashes with semantic structure. A randomly initialized and trainable collection of proxies \(P_{\theta} = \{\mathbf{p}_{\theta1}, \mathbf{p}_{\theta2},\dots, \mathbf{p}_{\theta N_{\text{cls}}}\}\) is used with a teacher hash \(\teachhash\) to compute a class-wise prediction \(\mathbf{p}_T = [\mathcal{S}(\mathbf{p}_{\theta1}, \teachhash), \mathcal{S}(\mathbf{p}_{\theta2}, \teachhash), \dots, \mathcal{S}(\mathbf{p}_{\theta N_{\text{cls}}}, \teachhash)]\), where \(N_{\text{cls}}\) is the number of classes. The class-wise prediction's similarity with the class label \(\c\) is then learned via a hash proxy loss
    \begin{align}\label{eq:HP_loss}
        \mathcal{L}_{\text{HP}}(\c, \mathbf{p}_T, \tau) \triangleq H\left(\frac{\c}{\|\c\|_1}, \text{Softmax}\left(\frac{\mathbf{p}_T}{\tau}\right)\right)
    \end{align}
    where \(H(\mathbf{u},\mathbf{v}) \triangleq -\sum_k u_k\log v_k\) is the cross entropy; softmax is computed along \(\mathbf{p}_T\); \(\|\!\cdot\!\|_1\) is the \(L_1\)-norm; and \(\tau\) is a temperature scaling hyperparameter~\cite{DeepHashDist}.

    The loss function for the DHD module is
    \begin{align}\label{eq:dhd_loss}
        \mathcal{L}_{\text{DHD}} & \triangleq \mathcal{L}_{\text{HP}} + \lambda_{\text{SdH}}\cdot{\mathcal{L}_{\text{SdH}}} + \lambda_{\text{bce-}Q}\cdot\mathcal{L}_{\text{bce-}Q}
    \end{align}
    where $\lambda_{\text{SdH}}$ and $\lambda_{\text{bce-}Q}$ are hyperparameters set to balance the training objectives~\cite{DeepHashDist}.
    
    We remark that deep hashes are not hashes in the cryptographic sense, as DHD hashes are intended for database retrieval, and thus, carry semantic information about the images. However, this feature allows us to align images in the semantic space via the cosine distance.

    \def\inferenceDHDmod{\DHDmodule^{\star}}
    \subsection{Training of DeepJSCC with DHD}\label{subsec:joint_train}
    \begin{figure}[t]
	    \centering
	    \includegraphics[width=0.82\linewidth]{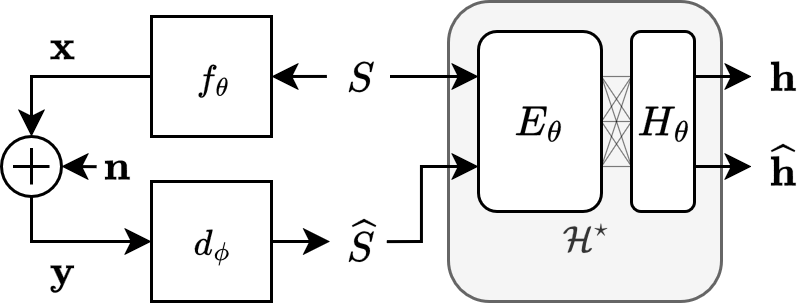}
	    \caption{System design for DeepJSCC-DHD with frozen \(\inferenceDHDmod\).}
	    \label{fig:system_design}
	\end{figure}

    We define a DeepJSCC scheme with a trainable, non-linear encoder \(f_\theta\) and decoder \(d_\phi\) as follows:
    \begin{align}
        f_\theta &: \mathbb{R}^{C \times H \times W} \rightarrow \mathbb{C}^{C_{\text{out}}/2\times H/4 \times W/4}, \\ 
        d_\phi &: \mathbb{C}^{C_{\text{out}}/2\times H/4 \times W/4} \rightarrow \mathbb{R}^{C \times H \times W}
    \end{align}
    where\\ 
    (i) $\theta$ and $\phi$ are trainable parameters of the encoder \(f_\theta\) and decoder \(d_\phi\), respectively,\\
    (ii) \(C_{\text{out}}\!\geq\!1\) is a hyperparameter, and\\
    (iii) \(k \triangleq{C_{\text{out}}/2\times H/4 \times W/4}\).

    We define our \emph{baselines} as (i) DF with DeepJSCC encoders and decoders \(f_{\theta,r+1}\) and \(d_{\phi,r+1}\) for \(r=[0,1,2,3]\); and (ii) QF with trained and frozen encoder and decoder \(f_{\theta^\star}\) and \(d_{\phi^\star}\). Both baselines are trained using \emph{only} MSE in the loss function. The QF setting's trained encoder-decoder pair is initialized with a trained pair from the DF setting when \(r\!=\!0\). We use the DeepJSCC architecture in~\cite{DJSCC-NOMA}, but modify it to use a single encoder-decoder pair without device embeddings.

    Our \emph{proposed} system first trains and \emph{freezes} a DHD module, denoted as \(\inferenceDHDmod\), by setting its parameters' learning rate \(lr\!=\!0\). This allows gradient computation and flow during backpropagation without updating the weights of the DHD module, which prevents hashes from collapsing to a trivial solution.
    
    Our proposed DeepJSCC encoder-decoder structures are identical to the baselines', except we train DeepJSCC with DHD to achieve semantic alignment between the source and reconstructed images \(\S\) and \(\widehat{\S}\). The objectives of minimizing \(\mathcal{L}_{\text{MSE}}\) and (\ref{eq:sdh_loss}), i.e., minimizing the pixel-wise error and simultaneously aligning the hash outputs \(\mathbf{h}=\DHDmodule(\S)\) and \(\widehat{\mathbf{h}}=\DHDmodule(\widehat{\S})\), provides the DeepJSCC module with semantic guidance that also improves the perceptual quality of \(\widehat{\S}\). We illustrate a proposed DF scenario for \(r\!=\!0\) in Fig.~\ref{fig:system_design}.
   
    Using MSE and (\ref{eq:sdh_loss}), define the loss function of the proposed system as 
    \begin{align}\label{eq:loss_MAIN}
        \mathcal{L}^{*} = \mathcal{L}_{\text{MSE}}(\S, \widehat{\S}) + \lambda\cdot\mathcal{L}_{\text{SdH}}(\mathbf{h}, \widehat{\mathbf{h}})
    \end{align}
    where \(\lambda\) is a hyperparameter to balance the objectives, and hashes \(\mathbf{h}\) and \(\widehat{\mathbf{h}}\) are outputs from \(\inferenceDHDmod\) with \(\S\) and \(\widehat{\S}\) as respective inputs. Note that we do not enforce any objectives for relay reconstructions \(\widetilde{\S}\) in either our system or the baselines.    
    
    \subsection{Experimental Setup}\label{sec:training}
    The dataset considered in this work is a subset of the NUS-WIDE dataset~\cite{NUSWIDE}, consisting of \(9,\!450\!:\!1,\!050\!:\!2,\!100\) training, validation, and test images with \(256\times256\) resolution (\(\S\!\in\!\mathbb{R}^{3 \times 256 \times 256}\), where exponent corresponds to \(C\times H \times W\)) and corresponding \(N_{\text{cls}}\!=\!21\) dimensional multi-hot encoded class labels \(\c\). We use a pre-trained ResNet50~\cite{ResNet50} as the encoder \(E_\theta\) with the feature dimension \(N_E\!=\!2048\). We set the hash length as \(N_H\!=\!64\) bits, and the remaining parameter values are assigned as default in~\cite{DeepHashDist}. The weights of \(H_\theta\) and \(E_\theta\) are then trained/fine-tuned as in~\cite{DeepHashDist} and frozen.

    We set the bandwidth ratio to \(\rho=\frac{1}{3}\), corresponding to \(C_{\text{out}}=32\) and \(k=\frac{CWH}{3} = 65,536\), and set \(\lambda=0.06\)~\cite{DeepJSCCandDHDmaster} in (\ref{eq:loss_MAIN}). We use the Adam~\cite{ADAM} optimizer with a \(lr=10^{-4}\) and a \verb|MultiplicativeLR|~\cite{PyTorch} scheduler that updates the learning rate as \(lr:=0.95lr\) at each epoch. To ensure fair comparisons between our proposed system and the baseline, they are trained and tested with identical hyperparameter settings and DeepJSCC architectures.
    
    For DF, we use mini-batch sizes \([20, 10, 8, 6]\) for training \(r=[0,1,2,3]\) relays, respectively (larger batch sizes cause memory overflows on NVIDIA Tesla V100 32GB GPUs). We train and test them at SNRs \([-5,-10,-15]\) dB, i.e., we train 12 DF setting models. We present our results in Section~\ref{sec:results-DF}.
    
    For the QF multi-hop setting, we rerun the validation set for trained DF models when \(r\!=\!0\) and collect channel outputs \(\y\). The collected outputs are used to compute \(2^{N_Qb}\) centers using the K-means algorithm on the collected \(\y\). We then rerun the test set while quantizing \(\y\) and dequantizing to \(\widehat{\y}\), from which we reconstruct a quantized image \(\widehat{\S}_Q=d_{\phi^*}(\widehat{\y})\). We test five levels of quantization, whose results are presented in Section~\ref{sec:results-QF}.
    
	\section{Experimental Results and Discussion}\label{sec:results}

    \subsection{DF Relaying}\label{sec:results-DF}
    
    \def\ourhighcol{blue}
    \def\ourmidcol{violet}
    \def\ourlowcol{darkgray}
    \def\imgscale{0.541}
    
    \begin{figure}[t]
        \centering

    \begin{tikzpicture}[scale=\imgscale]
        \begin{axis}[
        title={LPIPS for DF multi-hop relay},
        title style={yshift=-3pt},
        xlabel={Number of relays (\(r\))},
        xlabel style={yshift=5pt},
        ylabel={LPIPS},
        ylabel style={yshift=-6pt},
        xmin= -0.1, xmax=3.1,
        ymin= 0.1, ymax=0.57,
        xtick={0,1,2,3},
        ytick={0.10,0.20,0.30,0.40,0.50, 0.57},
        legend pos=south east,
        ymajorgrids=true,
        xmajorgrids=true,
        grid style=dashed,
        line width=1pt,
        mark options={solid}
        ]

        \addlegendimage{only marks,mark=square, color=black}
        \addlegendentry{Our DF}

        \addlegendimage{only marks, mark=triangle, color=black}
        \addlegendentry{DF Baseline}
        
        \addplot[
        dashed,
        color=\ourhighcol,
        mark=square,
        ] coordinates {
        (0,0.11038176)(1,0.180553123)(2,0.239513427)(3,0.252019614)
        }; 
        
        \addplot[
        dotted,
        color=\ourmidcol,
        mark=square,
        ] coordinates {
        (0,0.211272538)(1,0.305428475)(2,0.384727061)(3,0.411790013)
        }; 
        
        \addplot[
        dashdotted,
        color=\ourlowcol,
        mark=square,
        ] coordinates {
        (0,0.385191262)(1,0.485698432)(2,0.549591482)
        }; 

        \addplot[
        dashed,
        color=\ourhighcol,
        mark=triangle,
        ]
        coordinates {
        (0,0.122665204)(1,0.207732707)(2,0.248056501)(3,0.262799889)
        }; 
        \addplot[
        dotted,
        color=\ourmidcol,
        mark=triangle,
        ]
        coordinates {
        (0,0.261505723)(1,0.362200856)(2,0.39433378)(3,0.412250936)
        };
        \addplot[
        dashdotted,
        color=\ourlowcol,
        mark=triangle,
        ]
        coordinates {
        (0,0.440352529)(1,0.522717834)(2,0.551465154)(3,0.567913234)
        };
        
        \addplot[black]  coordinates{(2.5,0.225)} node[rotate=2] {\footnotesize SNR$=-5$~dB};
        \addplot[black]  coordinates{(2.5,0.375)} node[rotate=5] {\footnotesize SNR$=-10$~dB};
        \addplot[black]  coordinates{(2.5,0.54)} node[rotate=3] {\footnotesize SNR$=-15$~dB};
    \end{axis}
    \end{tikzpicture}
    \hfill
    \begin{tikzpicture}[scale=\imgscale]
        \begin{axis}[
        title={PSNR for DF multi-hop relay},
        title style={yshift=-3pt},
        xlabel={Number of relays (\(r\))},
        xlabel style={yshift=5pt},
        ylabel={PSNR [dB]},
        ylabel style={yshift=-3pt},
        xmin= -0.1, xmax=3.1,
        ymin=18.9, ymax=29,
        xtick={0,1,2,3},
        ytick={19, 21, 23, 25, 27, 29},
        legend pos=north east,
        ymajorgrids=true,
        xmajorgrids=true,
        grid style=dashed,
        line width=1pt,
        mark options={solid}
        ]
        \addlegendimage{only marks,mark=square, color=black}
        \addlegendentry{Our DF}

        \addlegendimage{only marks, mark=triangle, color=black}
        \addlegendentry{DF Baseline}

        \addplot[
        dashed,
        color=\ourhighcol,
        mark=square,
        ] coordinates {
        (0,27.63710976)(1,25.67710114)(2,24.5970993)(3,24.43899536)
        };
        \addplot[
        dotted,
        color=\ourmidcol,
        mark=square,
        ] coordinates {
        (0,24.64369202)(1,22.93259239)(2,21.82546425)(3,21.46220779)
        };
        \addplot[
        dashdotted,
        color=\ourlowcol,
        mark=square,
        ] coordinates {
        (0,21.01987839)(1,19.72975349)(2,18.96705627)
        }; 
        
        \addplot[
        dashed,
        color=\ourhighcol,
        mark=triangle,
        ]
        coordinates {
        (0,28.91051292)(1,27.27896881)(2,26.52289963)(3,26.30274582)
        }; 
        \addplot[
        dotted,
        color=\ourmidcol,
        mark=triangle,
        ]
        coordinates {
        (0,25.68112373)(1,24.4566288)(2,23.87874603)(3,23.63900948)
        };
        \addplot[
        dashdotted,
        color=\ourlowcol,
        mark=triangle,
        ]
        coordinates {
        (0,22.52273941)(1,21.57967758)(2,21.08730507)(3,20.7668972)
        };
        \addplot[black]  coordinates{(2,25)} node[pos=1,above] {\footnotesize SNR$=-5$~dB};
        \addplot[black]  coordinates{(2,22.5)} node[pos=1,above] {\footnotesize SNR$=-10$~dB};
        \addplot[black]  coordinates{(2,20)} node[pos=1,above] {\footnotesize SNR$=-15$~dB};
    \end{axis}
    \end{tikzpicture}
    \vspace{-20pt}
            \caption{DF multi-hop relay performance measured in LPIPS and PSNR. The line styles \{dashed, dotted dash-dotted\} belong to the SNRs \(\{-5,-10,-15\}\)~dB, respectively.}
        \label{fig:relay_results}
    \end{figure}

    The experimental results for the DF multi-hop relaying are presented in Fig.~\ref{fig:relay_results}. We note that a lower LPIPS score is better.
    We observe from Fig.~\ref{fig:relay_results} that PSNR and LPIPS performance decrease for both systems as the number of hops increases, as expected. Our results show that our system consistently reconstructs images with higher perceptual similarity compared to the baseline. LPIPS gain is more pronounced for lower SNRs, indicating that semantic clustering successfully aligns the DeepJSCC system toward reconstructing perceptually higher quality images. However, our system achieves a lower PSNR than the baseline, which is likely due to the reconstruction process sacrificing pixel-wise accuracy for the benefit of aligning semantic hashes. The difference in PSNR increases with the number of hops, which can be due to the proposed system maximizing the semantic alignment of the source and reconstructed image, imposing a constraint that leads to an increasing sacrifice of pixel fidelity while retaining semantic alignment performance.

    Training all architectures in \(-10\)~dB DF scenario using LPIPS with weight \(\alpha\in [0.01,1]\) as an additional loss term, we observe that for all \(\alpha\) our proposed architecture improves semantic alignment, measured by (\ref{eq:sdh_loss}), compared to the baseline. As \(\alpha\) increases, PSNR and LPIPS values of our architecture converge toward those of the baseline, with roughly equivalent PSNR and LPIPS values when training with \(\alpha \geq 0.6\). Thus, our architecture significantly improves semantic-alignment performance when trained with LPIPS, with similar perceptual quality to the baseline.
    
    \subsection{QF Relaying}\label{sec:results-QF}

    The experimental results for QF multi-hop relaying are depicted in Fig.~\ref{fig:quant_results}, where we have the parameters
    \begin{align}
        (N_Q, b) = \{(4, \frac{3}{4}), (2,1), (4, \frac{5}{4}), (2,\frac{3}{2}),(2,2)\}
    \end{align}
    in order from left to right, and \(k=65,536\). We observe that both systems improve performance in both LPIPS and PSNR with a higher quantization rate, as expected. In particular, at bpp\(=4\), both PSNR and LPIPS approach the respective values measured in the DF setting (\(r=0\)). Moreover, in high-noise regimes (i.e., high combined channel and quantization noise), our proposed scheme maintains stable semantic alignment after quantization. Even when a baseline trained with LPIPS achieves a lower LPIPS score, our method provides this additional capability, which could prove useful for semantically enabled security-oriented applications.

   The memory footprints of the full DeepJSCC module (both encoder and decoder) and DHD are \(89.2\)~MB and \(94.4\)~MB, respectively. We note that both models' computational complexities are the same during inference, as only the weights are affected by the different training methodologies while the architectures are the same. We will study the trade-off between architecture complexity and performance, as we estimated empirical time complexity as \(0.094\) and \(0.046\) seconds per image during training and testing, respectively.
    \begin{figure}[t]
        \centering      
        \begin{tikzpicture}[scale=\imgscale]
        \begin{axis}[
        title={LPIPS for QF multi-hop relay},
        title style={yshift=-3pt},
        xlabel={bits per pixel (bpp)},
        ylabel={LPIPS},
        xlabel style={yshift=5pt},
        ylabel style={yshift=-6pt},
        xmin=1.4, xmax=4.1,
        ymin=0.1, ymax=0.55,
        xtick={1.5,2, 2.5, 3, 3.5, 4},
        ytick={0.10,0.20,0.30,0.40,0.50, 0.55},
        legend pos=south west,
        ymajorgrids=true,
        xmajorgrids=true,
        grid style=dashed,
        line width=1pt,
        mark options={solid}
        ]
        \addlegendimage{only marks,mark=square, color=black}
        \addlegendentry{Our QF}

        \addlegendimage{only marks, mark=triangle, color=black}
        \addlegendentry{QF Baseline}
        
        \addplot[
        dashed,
        color=\ourhighcol,
        mark=square,
        ] coordinates {
        (1.5,0.33447369933128357)(2,0.24969932436943054)(2.5,0.17452481389045715)(3,0.1484721153974533)(4,0.1230015754699707)
        }; 
        \addplot[
        dotted,
        color=\ourmidcol,
        mark=square,
        ] coordinates {
        (1.5,0.39824792742729187)(2,0.3195776343345642)(2.5,0.26694419980049133)(3, 0.24542352557182312)(4,0.2264430820941925)
        };
        \addplot[
        dashdotted,
        color=\ourlowcol,
        mark=square,
        ] coordinates {
        (1.5, 0.4961511790752411) (2, 0.4616377055644989) (2.5, 0.4299261271953583) (3, 0.4201123118400574) (4, 0.39939093589782715)
        };
        
        \addplot[
        dashed,
        color=\ourhighcol,
        mark=triangle,
        ]
        coordinates {
        (1.5,0.3863624334335327)(2,0.2759701907634735)(2.5,0.16776973009109497)(3, 0.13553884625434875)(4,0.1159680187702179)
        }; 
        \addplot[
        dotted,
        color=\ourmidcol,
        mark=triangle,
        ]
        coordinates {
        (1.5,0.4037438631057739)(2,0.3264691233634949)(2.5,0.2887256443500519)(3, 0.27393442392349243)(4,0.26193755865097046)
        };
        \addplot[
        dashdotted,
        color=\ourlowcol,
        mark=triangle,
        ]
        coordinates {
        (1.5,0.5365508198738098) (2,0.4842970073223114) (2.5, 0.45412880182266235) (3, 0.44325393438339233) (4, 0.4408620595932007)
        };

        \addplot[black]  coordinates{(3.5,0.15)} node[rotate=-7] {\footnotesize SNR$=-5$~dB};
        \addplot[black]  coordinates{(3.5,0.28)} node[rotate=-4] {\footnotesize SNR$=-10$~dB};
        \addplot[black]  coordinates{(3.5,0.46)} node[rotate=0] {\footnotesize SNR$=-15$~dB};
    \end{axis}
    \end{tikzpicture}
    \hfill
    \begin{tikzpicture}[scale=\imgscale]
        \begin{axis}[
        title={PSNR for QF multi-hop relay},
        title style={yshift=-3pt},
        xlabel={bits per pixel (bpp)},
        xlabel style={yshift=5pt},
        ylabel={PSNR [dB]},
        ylabel style={yshift=-3pt},
        xmin=1.4, xmax=4.1,
        ymin=16, ymax=27.5,
        xtick={1.5,2, 2.5, 3, 3.5, 4},
        ytick={17, 19, 21, 23, 25, 27},
        legend pos=north west,
        ymajorgrids=true,
        xmajorgrids=true,
        grid style=dashed,
        line width=1pt,
        mark options={solid}
        ]
        \addlegendimage{only marks,mark=square, color=black}
        \addlegendentry{Our QF}

        \addlegendimage{only marks, mark=triangle, color=black}
        \addlegendentry{QF Baseline}
        
        \addplot[
        dashed,
        color=\ourhighcol,
        mark=square,
        ] coordinates {
        (1.5, 17.59598159790039) (2, 19.389860153198242) (2.5, 22.206796646118164) (3, 23.88382911682129) (4, 26.062679290771484)
        };
        \addplot[
        dotted,
        color=\ourmidcol,
        mark=square,
        ] coordinates {
        (1.5, 17.373512268066406) (2, 19.582862854003906) (2.5, 21.498477935791016) (3, 22.691024780273438) (4, 23.863218307495117)
        };
        \addplot[
        dashdotted,
        color=\ourlowcol,
        mark=square,
        ] coordinates {
        (1.5, 16.63395881652832) (2, 18.070165634155273) (2.5, 19.31346321105957) (3, 19.768524169921875) (4, 20.535140991210938)
        };
        
        \addplot[
        dashed,
        color=\ourhighcol,
        mark=triangle,
        ]
        coordinates {
        (1.5, 16.535253524780273) (2, 19.1673526763916) (2.5, 22.190160751342773) (3,  24.181703567504883) (4,  26.868104934692383)
        }; 
        \addplot[
        dotted,
        color=\ourmidcol,
        mark=triangle,
        ]
        coordinates {
        (1.5, 17.443334579467773) (2, 19.671478271484375) (2.5, 21.963998794555664) (3, 23.3079891204834) (4, 24.750070571899414)
        };
        \addplot[
        dashdotted,
        color=\ourlowcol,
        mark=triangle,
        ]
        coordinates {
        (1.5, 17.68158721923828) (2, 19.406417846679688) (2.5, 20.70859146118164) (3, 21.368635177612305) (4, 22.042613983154297)
        };

        \addplot[black]  coordinates{(3.5,26)} node[rotate=26] {\footnotesize SNR$=-5$~dB};
        \addplot[black]  coordinates{(3.5,22.8)} node[rotate=13] {\footnotesize SNR$=-10$~dB};
        \addplot[black]  coordinates{(3.5,20.5)} node[rotate=8] {\footnotesize SNR$=-15$~dB};
    \end{axis}
    \end{tikzpicture}
    \vspace{-20pt}
          \caption{QF multi-hop relay performance measured in LPIPS and PSNR. The line styles \{dashed, dotted, dash-dotted\} belong to the SNRs \(\{-5,-10,-15\}\)~dB, respectively.}
          \label{fig:quant_results}
    \end{figure}
\section{Conclusion}\label{sec:conclusion}
    We demonstrated that our multi-hop DeepJSCC-DHD scheme significantly improves semantic alignment between the perceptual quality of the reconstructed images, measured by LPIPS, for both DF and QF relay settings by leveraging semantic clustering via a trained DHD module. Moreover, our experimental results showed that semantic alignment in our scheme remains robust also to quantization effects. Therefore, the proposed multi-hop DeepJSCC-DHD scheme adds a semantic alignment capability to DeepJSCC, complementing perception-oriented training and enabling secure authentication-oriented DeepJSCC applications. 
    
    \newpage
	\section{Acknowledgement}
    This work has been supported by the ZENITH Research and Leadership Career Development Fund under Grant ID23.01, the Swedish Foundation for Strategic Research (SSF) under Grant ID24-0087, the SNS JU project 6G-GOALS under the EU’s Horizon program (Grant Agreement No. 101139232), and the German Federal Ministry of Research, Technology and Space (BMFTR) 6GEM+ Transfer Hub under Grant 16KIS2412. The computations and data handling were enabled by resources provided by the National Supercomputer Centre (NSC), funded by Linköping University.
	
	\bibliographystyle{IEEEbib}
	\bibliography{refs}

\end{document}